\documentclass[twocolumn,showpacs,preprintnumbers,amsmath,amssymb]{revtex4}
\usepackage{graphicx}
\begin{document}
\title{Phonon-assisted  transport through double-dot Aharonov-Bohm interferometer in the Kondo regime}
\author{P. P. Flork\'{o}w, D. Krychowski and S. Lipi\'nski}
\affiliation{%
Institute of Molecular Physics, Polish Academy of Sciences\\M. Smoluchowskiego 17,
60-179 Pozna\'{n}, Poland
}%
\date{\today}
\begin{abstract}
The effect of electron-phonon coupling on transport through a pair of strongly correlated quantum dots embedded in the Aharonov-Bohm ring is considered in the mean field slave boson Kotliar-Ruckenstein approach. It is shown that coupling with phonons opens transport gap in the region of double occupancy. Low-bias conductance and thermopower provide information on electron-phonon coupling strength.\end{abstract}
\maketitle

\section{Introduction}
There is currently a great interest in the interplay of strong correlations and interference in multiply connected geometries with embedded quantum dots or molecular ring systems \cite{ref1, ref2}. One of the ways to modify the interference conditions is to use the magnetic field (Aharonov-Bohm (AB) oscillations \cite{ref2}).  Recently much attention is also paid to the effects of local vibrations, because electronic and phonon energies in nanoscopic systems can become of the same order of magnitude generating scenarios, where novel effects may emerge and consequently new paths of device functionality appear. The present paper is devoted to the analysis of the impact of electron-phonon coupling on magnetotransport in the Kondo range. It is shown that due to the  capacitive coupling of the dots an information about electron-phonon coupling of the subsystem  is  transferred  to the subsystem not directly  coupled to vibrations. This  manifests itself in the structure of   low voltage conductance and thermopower. Asymmetric coupling of phonons to different dots strongly influences interference conditions and this in turn reflects in sharpening of AB oscillations.

\section{Model and formalism}
We study a pair of capacitively coupled dots embedded in AB ring coupled with local vibration modes described by Anderson-Holstein model with equal intra and interdot Coulomb interactions:
\begin{figure}
\centering
\includegraphics[width=4.5cm]{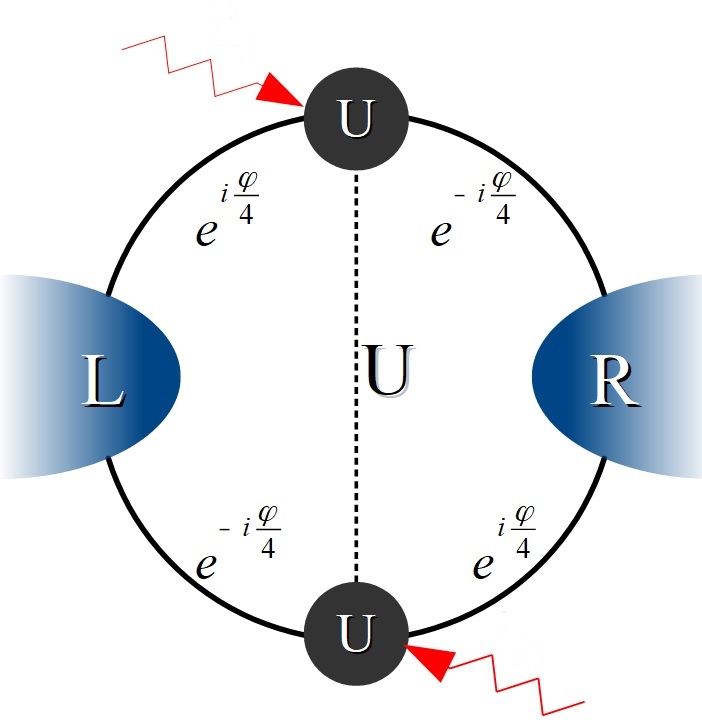}
\caption{Schematic view of AB ring with embedded dots coupled to local vibrations.}
\label{fig1}
\end{figure}
\begin{eqnarray}
&&\mathcal{H} = \sum_{k \alpha \sigma} E_{k \alpha \sigma} c_{k \alpha \sigma}^{\dagger} c_{k \alpha \sigma} + \sum_{i \sigma} E_{i \sigma}d_{i \sigma}^{\dagger} d_{i \sigma} +\nonumber\\&&U\sum_{\sigma \bar{\sigma}}n_{1\sigma}n_{2\bar{\sigma}}+U\sum_{i} n_{i \uparrow}n_{i \downarrow}+\\&&\sum_{k \alpha i \sigma} t (e^{i \frac{\varphi}{4}} c_{k \alpha \sigma}^{\dagger} d_{i\sigma} + e^{-i \frac{\varphi}{4}} c_{k \alpha \sigma}^{\dagger}d_{i\sigma} + h.c)+\nonumber\\&&\omega_{0}\sum_{i}b_{i}^{\dagger}b_{i} + \sum_{i \sigma}\lambda_{i}(b_{i}+b_{i}^{\dagger})d_{i\sigma}^{\dagger}d_{i\sigma}\nonumber,
\end{eqnarray}
where $c_{k \alpha \sigma}^{\dagger}$ creates an electron state in left(right) metallic electrode $\alpha = L(R)$. $n_{i \sigma} = d_{i \sigma}^{\dagger}d_{i \sigma}$ is occupation number of dot i, $E_{k \alpha \sigma}$($E_{ i \sigma}$) denotes energy of electrons in lead (dot). Intra- and intercoulomb interactions are parametrized by U and $t$ describes hopping between leads and dots, which is farther modified by Peierls phase factors $e^{\pm i \frac{\varphi}{4}}$, where $\varphi=2\pi \frac{\phi}{\phi_0}$, $\phi$ is the magnetic field flux and $\phi_0$ is magnetic flux quantum. $b_{i}^{\dagger}$  creates phonon with vibrational frequency $\omega_0$ coupled to electrons on the dot i with strength $\lambda_i$. We set $|e| = |g| = |B| = |k_B| = |h| = 1$. Assuming  strong electron-phonon (e-ph) interaction regime ($\lambda_i>> t$) we eliminate the linear e-ph coupling terms by the Lang-Frisov-type (LF) unitary transformation describing formation of local polarons \cite{ref3}. Transformed hamiltonian takes the form $\mathcal{\tilde{H}}=e^{S}He^{-S}$ with $S=S_1 + S_2$, $S_i=\frac{\lambda_i}{\omega_0}\sum_{\sigma}d_{i\sigma}^{\dagger}d_{i\sigma}(b_i^{\dagger}-b_i)$. The relevant parameters of (1) become renormalized: $\tilde{E}_{i\sigma}=E_{d}-\frac{\lambda_{i}^{2}}{\omega_0}$, $\tilde{U}_{i}=U_{i}-2\frac{\lambda_{i}^{2}}{\omega_0}$. Assuming $t<<\lambda_i$, one can adopt independent boson model \cite{ref3} and the phonon induced suppression of tunneling then reads $\tilde{t}_{i\sigma}=t_{i\sigma}e^{-\frac{\lambda_{i}^2}{\omega_0^2}}$. LF transformation causes the electron and phonon subsystems to decouple, therefore the corresponding averages can be calculated independently. The Fourier transforms of the electron lesser Green's functions for $T=0$ read: $G_{ij}^{<}(E)=\sum_{n=-\infty}^{\infty}L_n^{ij}(\lambda_1,\lambda_2, \omega_0)\tilde{G}_{ij}^{<}(E+n\omega_0)$, with $L_n^{ij}=e^{-(\frac{\lambda_i}{\omega_0})^2}(\frac{\lambda_i}{\omega_0})^{2n}(\frac{1}{n!})$ and $L_n^{12}=L_n^{21}=\sqrt{L_n^{11}L_n^{22}}$. To discuss correlation effects, we use finite U slave boson mean field approach (SBMFA) of Kotliar and Ruckenstein \cite{ref4}. A set of 16 auxiliary bosons are introduced acting as projection operators onto empty, fully occupied double dot system and different double and triple occupied states.  By imposing constraints on the completeness of the states and charge conservation one establishes the physically  meaningful sector of the Hilbert space. These constraints can be enforced by introducing corresponding Lagrange multipliers \cite{ref4}. In MFA the SB operators are replaced by their expectation values and formally the problem is reduced to the effective free electron model with renormalized hopping integrals and dot energies.

Using the Onsager relations for the particle current and the heat flux, one yields the formulas for the electric conductance $\mathcal{G}$ and the thermopower $\mathcal{S}$ in the forms: $\mathcal{G}(V)=-\frac{e^2}{T}L_0$ and $\mathcal{S}=-\frac{1}{e T}\frac{L_1}{L_0}$ \cite{refZ}, where the linear response coefficients are given by $L_{0(1)}=\frac{2T}{h}\sum_{\alpha}\int(E-\mu_{\alpha})^{0(1)}(\frac{\partial f_{\alpha}}{\partial \mu_{\alpha}})_{T}\mathcal{T}(E, \omega_0, \varphi)dE$, where $f_{\alpha}$ denotes Fermi-Dirac distribution functions and $\mathcal{T}(E, \omega_0, \varphi) = \mathcal{T}_1 + \mathcal{T}_2$ is the total transmission through the QD system.

\section{Results and discussion}
In presentation of numerical results we use relative energy units choosing $D/50$ as the unit, where $D$ is the electron bandwidth. The calculations were carried out in the strong correlation regime assuming Coulomb parameter $U=3$ and coupling to the leads $\Gamma_L=\Gamma_R=\frac{\pi t^2}{D}=0.05$. Phonon energy was taken as $\omega_0=0.5$. Figure 2a compares densities of states (DOS) of considered double dot system (DQD) (Fig. 1) for the case when two local Einstein phonon modes $\omega_0$ are separately coupled to each of the dots with equal strength with the case when local vibrations are absent. The DQD system in these cases is fully symmetric, both dots are equally connected to the common electrodes and in the latter case also equally coupled to local vibrations. For $\lambda_1=\lambda_2=0$ Kondo-Dicke effect is seen \cite{ref5, ref6}, which is a manifestation of an interplay of interference introduced by indirect coupling of the dots via common electrode (Dicke effect) and Kondo correlations. As a result the antibonding Kondo resonance localizes at the Fermi level for $E_d=-U/2$ and becomes extremely sharp (Dirac $\delta$ peak). It does not contribute to the linear conductance (dark state), whereas the bonding (broad resonance) contributes to the linear transmission. Inset of Fig. 2a displays the same densities of states but on a wider energy scale. Side peaks replicating Kondo-Dicke resonances are seen for energies that are multiples of $\omega_0$. Although they are observed in DOS again they do not contribute to transmission (not presented).

\begin{figure}
\centering
\includegraphics[width=4cm]{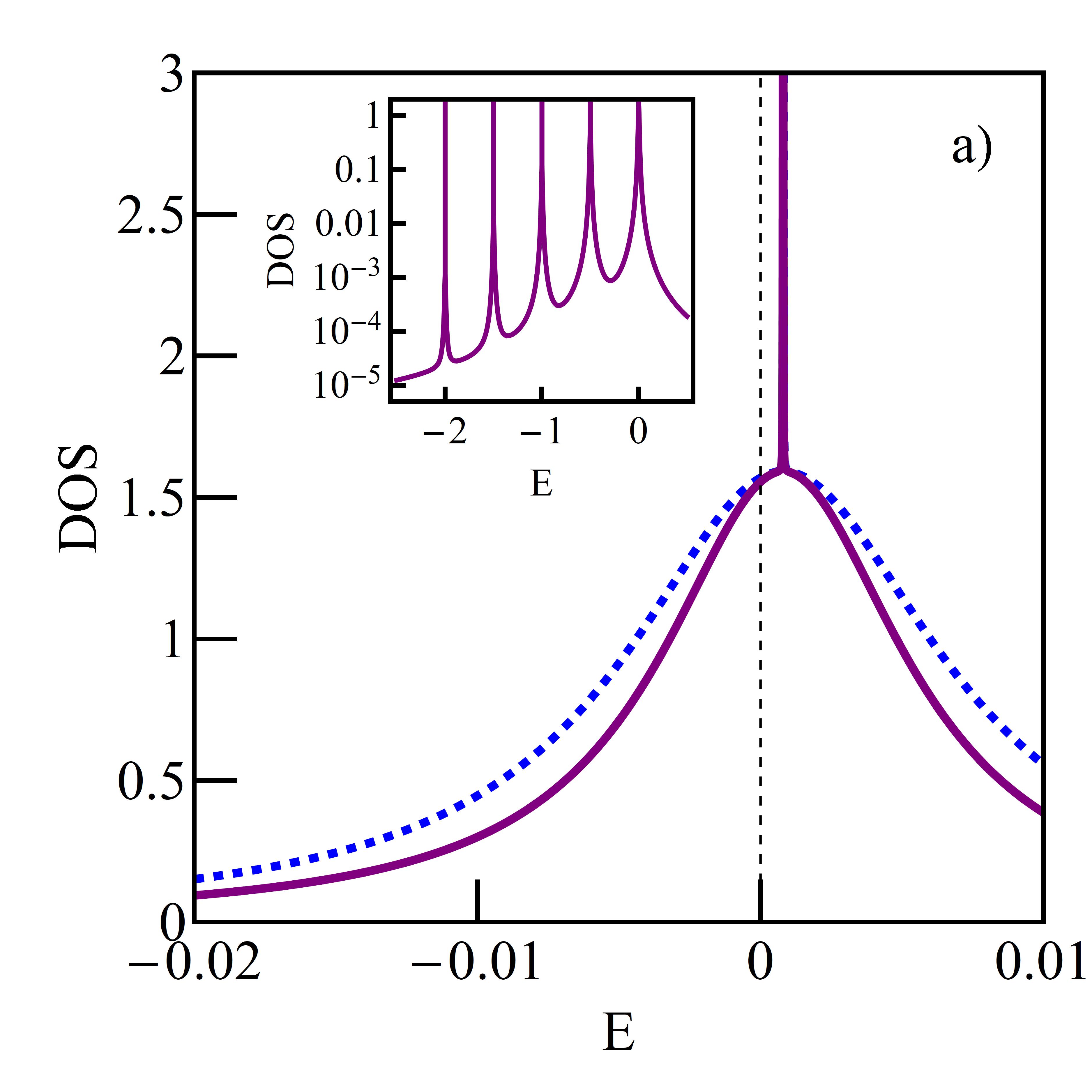}
\includegraphics[width=4cm]{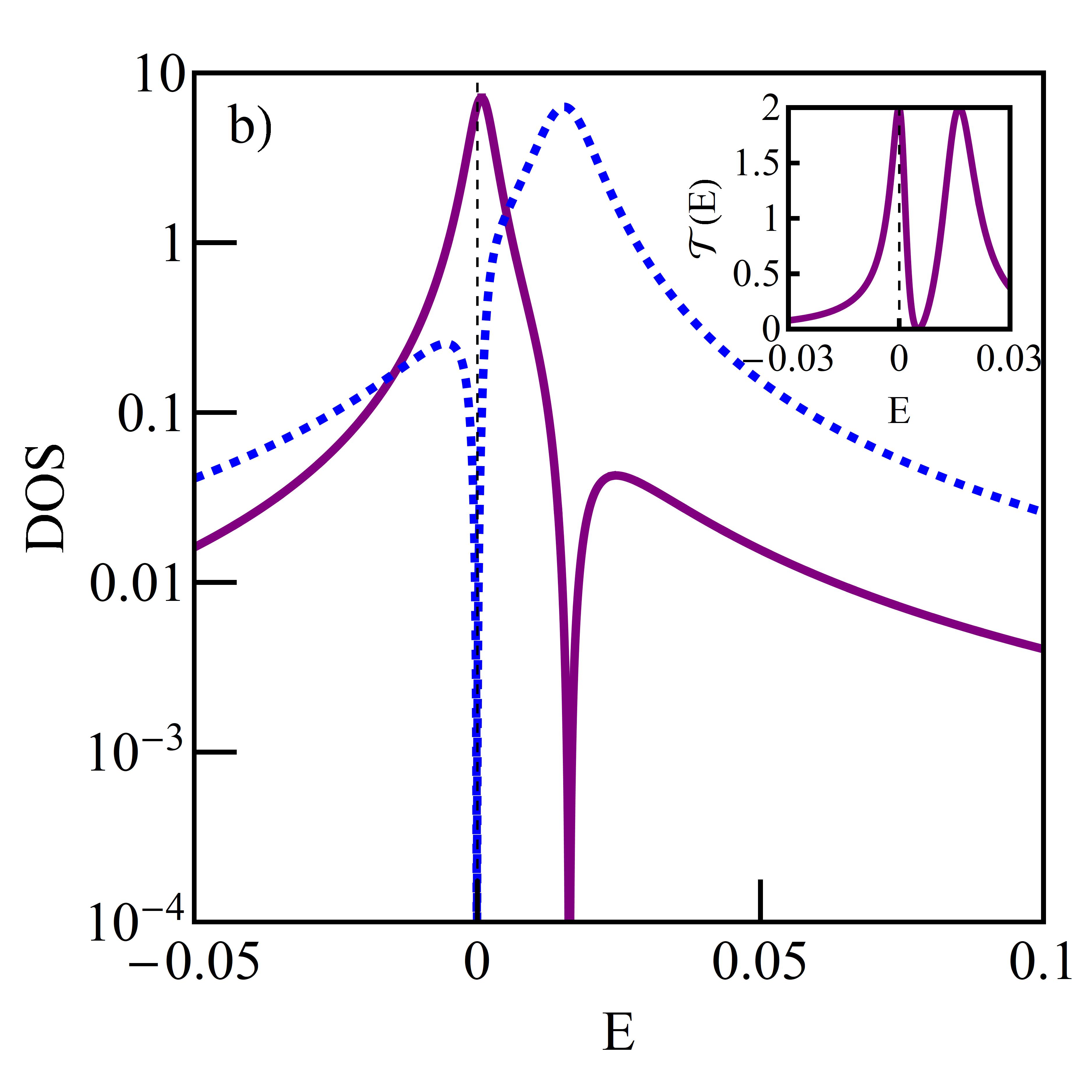}\\
\includegraphics[width=4cm]{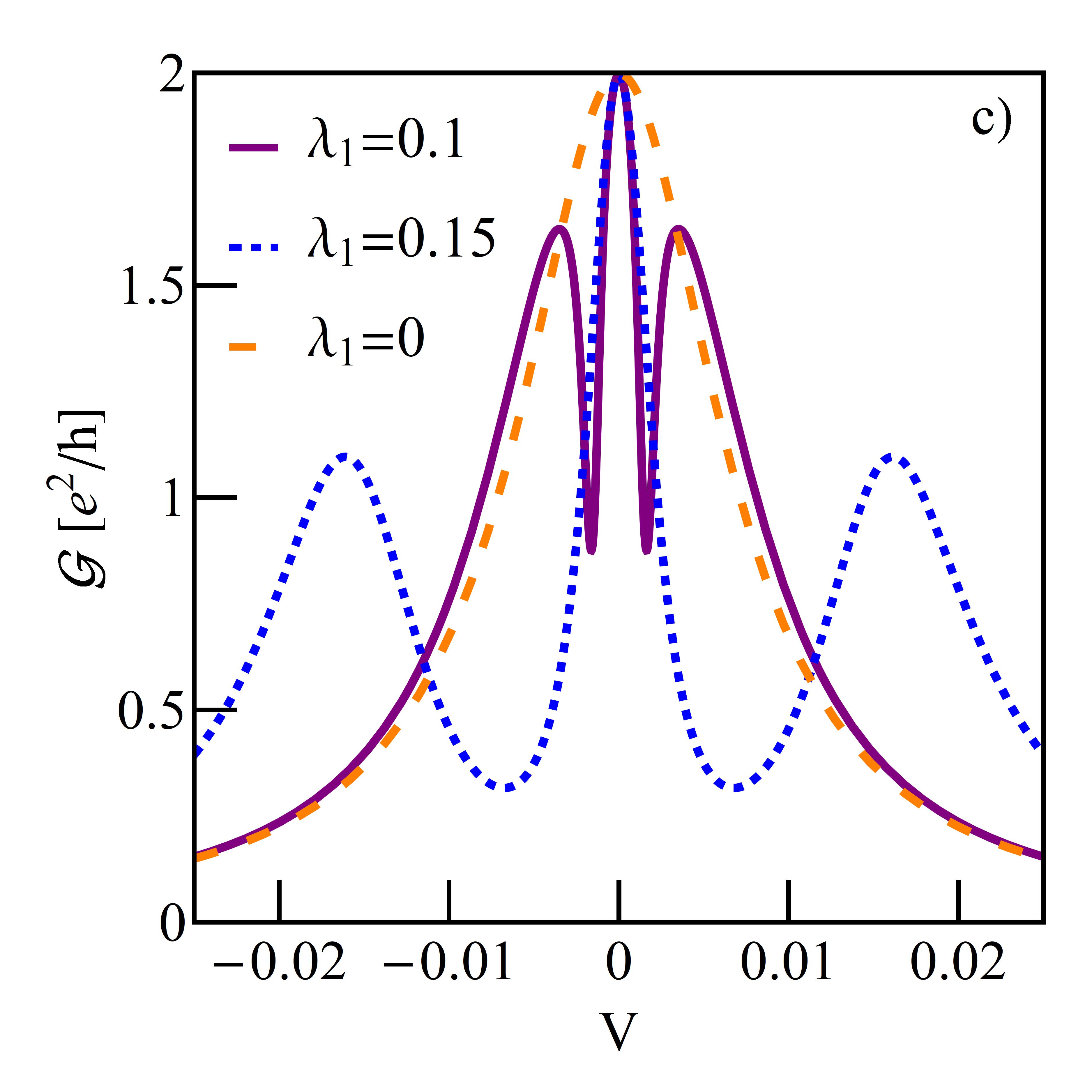}
\includegraphics[width=4cm]{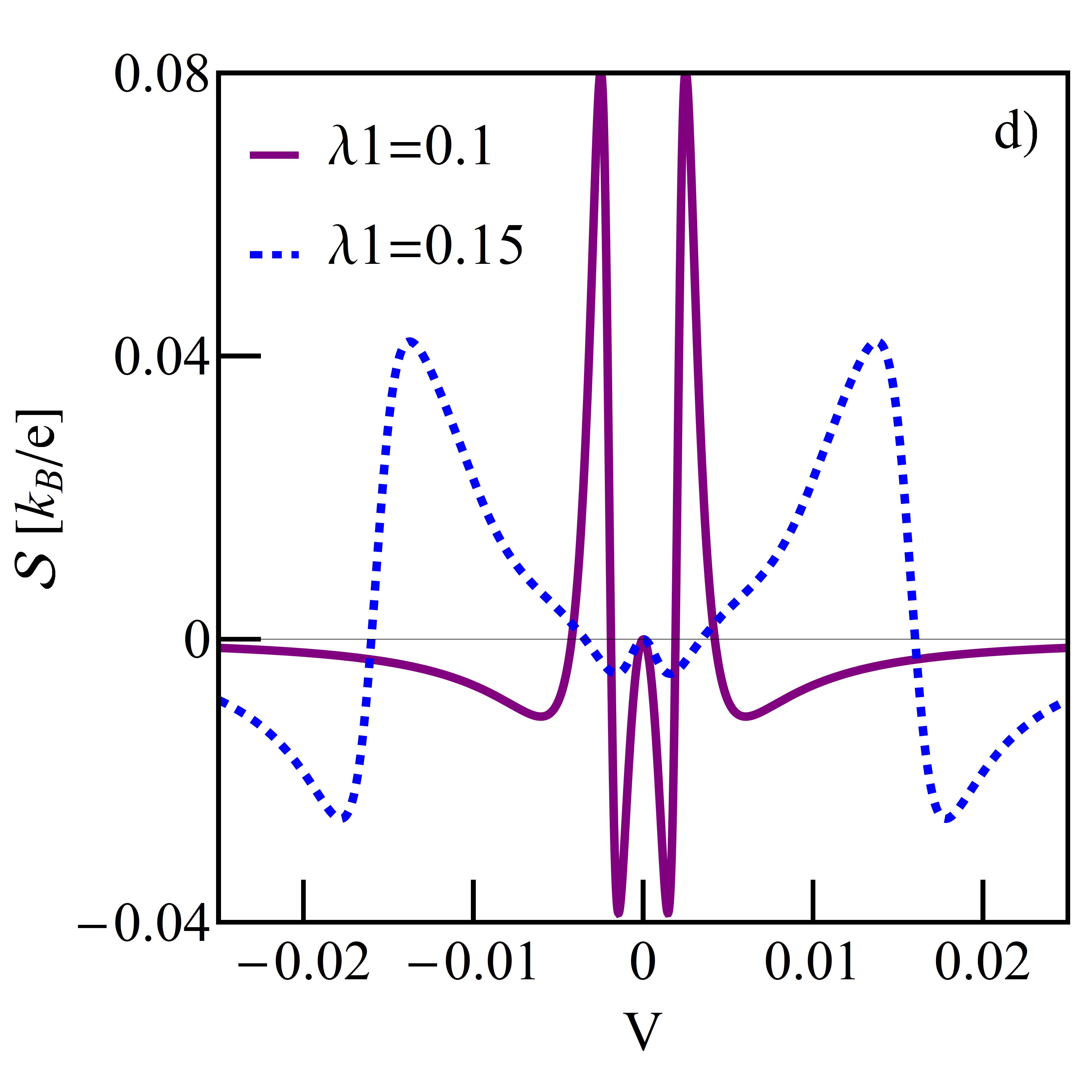}
\caption{(a) Total DOS of DQD ring for the case when each of the dot is equally and separately coupled to local phonon $\lambda_1 =\lambda_2 = 0.3$. Inset shows DOS in the broader energy range. (b) Partial densities of states for the case when only one of the dots (QD1) is coupled with phonon ($\lambda_1=0.15$). QD1 (solid line) and QD2 (dashed line), inset presents total low-energy transmission. (c) Total low-bias differential conductance of DQD ring and thermopower (d) for the case when only one of the dots is coupled to phonon. Thermopower was calculated for $T=10^{-6}$.}
\label{fig1}
\end{figure}

Fig. 2b illustrates the case when single local phonon mode is coupled to only one of the dots. The energies of bonding and antibonding states differ in this case due to phonon induced renormalization. We postpone the analysis of phonon assisted satellite peaks and limit ourselves only to the energy range close to the Fermi level ($E << \omega_0$). The presented DOSs have Kondo-Fano like shapes with maxima and dips located at the energies  of bonding or antibonding states. They correspond to constructive or destructive interference respectively. Inset presents the low energy transmission, resonance reaching the peak at the Fermi level comes predominantly from the many-body state of the dot, which is directly coupled to the local phonon (hereafter we will label this dot as dot 1), whereas the shifted resonance stems from the dot with no direct coupling to vibrations (dot 2). The manifestations of the discussed transmission in the linear conductance and thermopower for different values of e-ph couplings are shown on Figs. 2c, d. The central peak of conductance reaching the unitary limit of $2 e^2/h$ at zero bias describes transport through dot 1, whereas the satellites symmetrically located for positive and negative bias characterize low-voltage transport through the dot 2.  Important observation is that information about coupling of one of the dots with vibrations is also visible in ultra low voltage transport through another dot that does not directly interact with the lattice. This fact is also evident in thermopower in rapid changes of its sign at low voltages. Satellite positions of low-bias conductance ($\lambda_1^2/\omega_0$) carry the information about the strength of e-ph coupling.

\begin{figure}
\centering
\includegraphics[width=4cm]{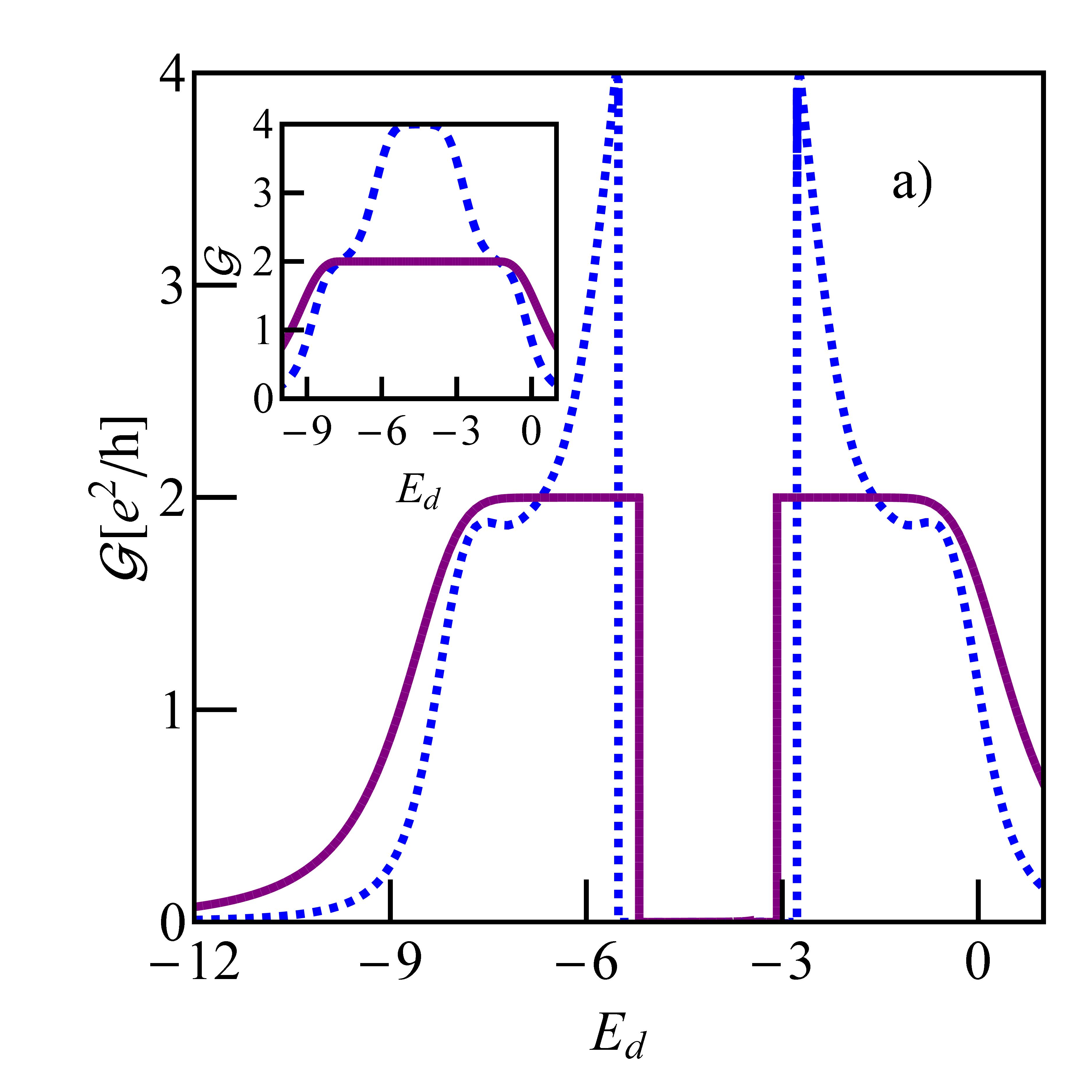}
\includegraphics[width=4.1cm]{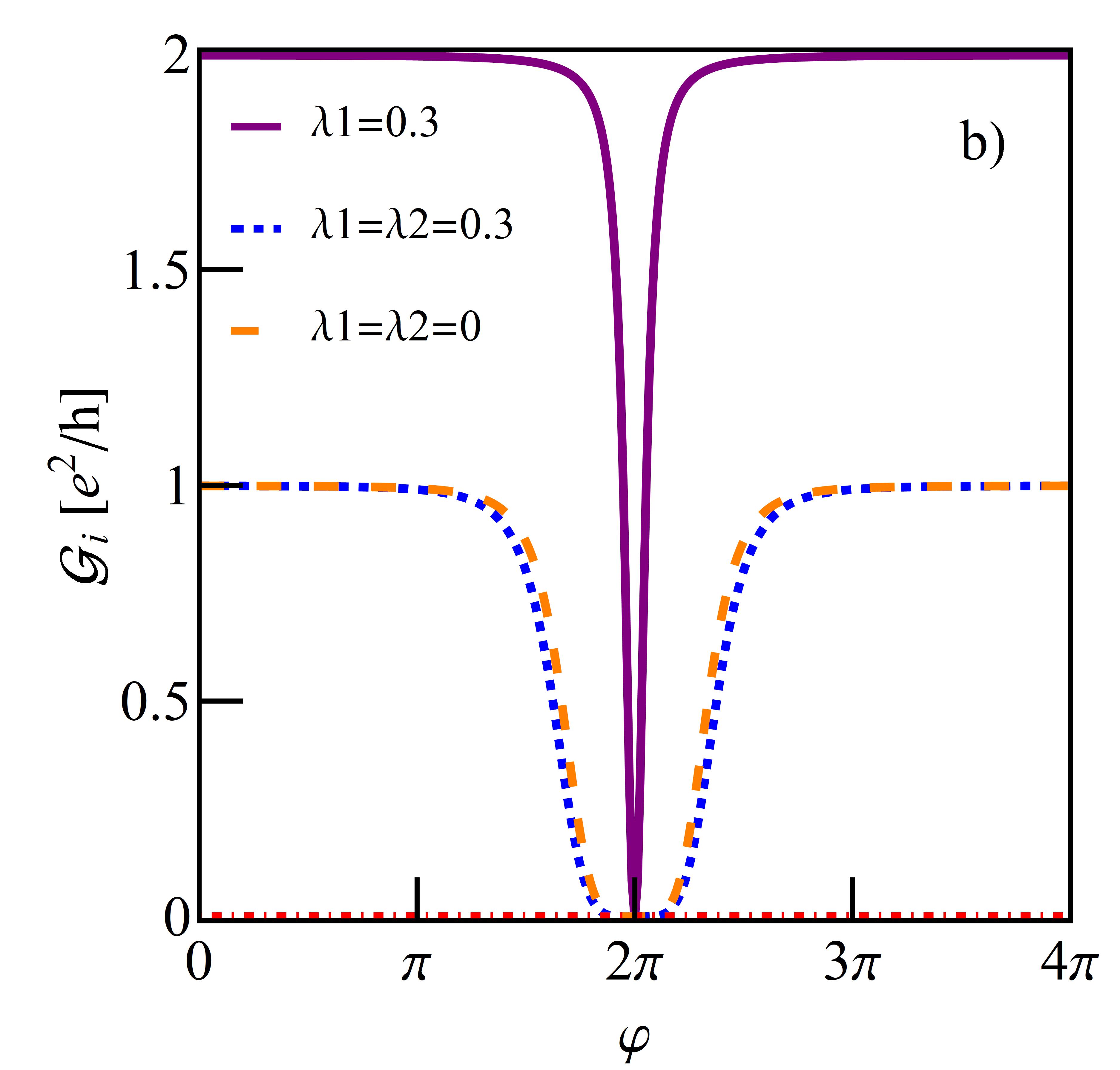}
\caption{ (a) Gate dependence of total conductance of DQD ring ($\lambda_{1}=\lambda_{2}=0.3$)(two level 2LSU(2) symmetry)(solid line) compared with similar dependence for parallel capacitively coupled DQD (SU(4))(dotted line). Inset shows the corresponding conductances for ($\lambda_1=\lambda_2=0$). (b) Aharonov-Bohm oscillations on QD1 for different e-ph couplings.}
\label{fig1}
\end{figure}

Fig. 3a presents the influence of phonons on gate dependence of conductance of DQD for equal e-ph couplings ($\lambda_1=\lambda_2$). In the region of double occupancy ($N \approx 2$) Kondo effect is destroyed and system transforms to charge ordered state ($N_1=2, N_2=0$) or ($N_1=0, N_2=2$) and blocking of conductance is observed \cite{ref7}. The double occupancy of single dot is preferred due to phonon induced weakening of intradot effective Coulomb interactions. Figure 3b illustrates the impact of e-ph coupling on Aharonov-Bohm oscillations presented for $Ed=-U/2$ $(N \approx 1)$. For the fully symmetric case of equal couplings of separate identical local phonons to the dots only the minor changes are observed. It is a consequence of the fact that for $N\approx 1$ the phonon induced modifications of interference conditions and correlations result mainly from the phonon suppression of coupling with the leads. Phonon induced shifts of dot energies and effective Coulomb interactions are only secondary in this case. The latter two effects might be of importance for systems, which are in the mixed valence state in the absence of phonons or for which a very strong e-ph coupling moves them towards this region. In this case significant modification of AB magnetoconductance introduced by phonons is expected. For asymmetric system with $\lambda_1\neq 0$ and $\lambda_2=0$ much steeper drop of conductance is observed around $\phi=2\pi$, because interference conditions are more strongly affected in this case than for symmetric coupling. It opens the possibility of influencing the switching capabilities of AB rings by phonons.

Summarizing, the present paper shows, that low bias transport characteristics inform about the strength of electron-phonon interaction, providing the unperturbed system is fully symmetric.  Above the critical value of e-ph  coupling transport gap opens in the region of double occupancy. Phonons strongly effect A-B oscillations for systems with phonons asymmetrically coupled to different dots.

\end{document}